\documentclass[aps,prd,showkeys,showpacs,amssymb,cite,
amsfonts,epsf,preprintnumbers,nofootinbib,superscriptaddress]{revtex4}

\usepackage[dvips]{graphicx}
\usepackage{bm,latexsym,amsmath,amssymb,amsfonts}
\usepackage[usenames,dvipsnames]{color}
\usepackage[colorlinks=true,linkcolor=blue]{hyperref}
\usepackage{color}
\usepackage{soul}
\newcommand{\be}[1]{\begin{equation} \label{#1}}
\newcommand{\ee}{\end{equation}}
\newcommand{\bea}{\begin{eqnarray}}
\newcommand{\eea}{\end{eqnarray}}
\newcommand{\ba}{\begin{array}}
\newcommand{\ea}{\end{array}}

\newcommand{\csch}{\mbox{csch}}
\newcommand{\arctanh}{\mbox{tanh}^{\mbox{-}1}}

\newcommand{\CG}{\mathcal{C}_\G}
\newcommand{\cZ}{\mathcal{Z}}
\newcommand{\G}{\mathcal{G}}
\begin{document}
\title{ %\tcr{\st{Gravity on Relativistic} 
Matter Equation of State in General Relativity }
\author{Hyeong-Chan Kim}
\email{hyeongchan@gmail.com}
\affiliation{Department of Physics, North Carolina State University, Raleigh, NC 27695-8202}
\affiliation{School of Liberal Arts and Sciences, Korea National University of Transportation, Chungju 380-702, Korea} 
\author{Chueng-Ryong Ji}
\email{crji@ncsu.edu}
\affiliation{Department of Physics, North Carolina State University, Raleigh, NC 27695-8202} 
\begin{abstract}
We study how a strong gravity affects the equation of state of matters. 
For this purpose, we employ a canonical ensemble of classical monoatomic ideal gas inside a box
in a Rindler spacetime. 
The total energy decreases monotonically with the increase of the external gravity representing its attractiveness. 
It is however bounded below, which is different from that of the Newtonian gravity case.   
As for the entropy, it decreases with the external gravity in the Newtonian regime.
However, in the presence of strong gravity or ultra-relativistic high temperature, the entropy increases with the gravity.
This result can be a resolution of the negative entropy problem of the ideal gas in the Newtonian gravity. 
In the presence of strong gravity, the bottom of the box is very close to the event horizon of the Rindler spacetime mimicking a blackhole
and the gas behaves as if it is on an effective two dimensional surface located at the bottom of the box.
Investigating the equation of state in the strong gravity regime, the temperature of the system is found to be not a free parameter but to approach 
a fixed value proportional to the external gravity, which is reminiscent of the Unruh temperature. 
\end{abstract}
\maketitle

%\pagecolor[rgb]{0.90,1.00,0.90}
% \begin{itemize}
%\item \colorbox{Yellow}{\color{Blue} How }
%\item The 
%\end{itemize}
%\pagecolor[rgb]{0.90,1.00,0.90}
%\pagecolor{black}
%----------------------------------------------
%\color{white}
\section{Introduction}
An equation of state (EOS) is a thermodynamic relation describing the state of matter under a given set of physical conditions. 
The well-known EOS for an ideal gas is
\be{EoS}
PV= N k_B T,
\ee
where $P$, $V$, $N$, $k_B$, and $T$ represents the pressure, volume, total number of particles, Boltzmann constant, and temperature for the system, respectively.
Given a gravity theory, this equation is not sufficient to construct star structures until an additional constraint is given relating the thermodynamic variables. 
The adiabaticity or the isentropic condition is usually used for this purpose.
The relation accompanying this constraint is also called the EOS. 
Various EOSs are used to describe systems with strong gravity such as the sun, white dwarf stars, neutron stars~\cite{Lattimer:2012nd} and early universe~\cite{cosmology}.
A typical example is the polytropic EOS, $P = k_{\rm poly} 
\rho^\gamma$, where $\rho$ is the energy density and the associated effective constants describing the system are denoted by $k_{\rm poly}$ and $\gamma$. 

Conventionally in general relativity, an EOS for a system is obtained in flat spacetime.
So-obtained EOS is used even for the system with gravity because a freely falling frame is still locally flat. 
Thus, the pressure and the density in the frame can be obtained from the same way as those in the flat spacetime, which are used as the input values of stress-tensor in a locally orthonormal frame. 
Although this prescription works well for most cases, we should examine whether it works or not for extreme cases when a very strong gravity or high curvature is present.
In Ref.~\cite{Kim:2016txr}, the EOS of matters obtained without gravity is shown to be credible for most astrophysical systems even in the presence of a Newtonian gravity.
However, it was also shown that there exist a few important situations in which the strong gravity over its average kinetic energy affects the EOS.
Examples are an isothermal self-gravitating system and a thin-shell of matters in extreme gravity.   
As a related work, the instability associated with a modification of the relation between pressure and volume was studied for a self-gravitating spherical mass 
of isothermal ideal gas~\cite{Bonnor,Lombardi:2001ms}  and for the nuclear matter in neutron stars in the context of a solitonic Skyrme model~\cite{adam}.
A possible modification of local EOS due to curvature effect was given~\cite{Kim:2013nna} in the context of the Eddington-inspired Born-Infeld gravity~\cite{Banados:2010ix,Pani:2012qd}.

To be concrete, a system in strong gravitational field requires a general relativistic treatment. 
For this purpose, we consider a system of monoatomic gas in Rindler spacetime, which is a general relativistic description of constant gravity.
Even though the work is done on a specific metric, the results are robust to general situations because every spacetime can be written as a Rindler metric locally~\cite{Jacobson:1995ab}.
The Rindler metric is given by
\begin{equation}\label{metric}
ds^2 = g_{\mu\nu}dx^\mu dx^\nu=  -\big(1+gz\big)^2dt^2 + dx^2+ dy^2+ dz^2,
\end{equation}
where $g$ represents the gravity acting on a particle at $z=0$.
For notational simplicity, we set $c=1$. 
To restore $c$ in this work, one may use the prescription $g \to g/c^2$, $t \to ct$, $v \to v/c$, $\mu_0 \to \mu_0c^2$, $p/\mu_0\to p/\mu_0c$, and
$\hbar/\mu_0 \to \hbar/\mu_0 c$. 
Occasionally, we restore $c$ to show the dimension as explicit as possible.
The surface $z= -g^{-1}$ corresponds to an event horizon where the metric can be extendible beyond the horizon to give a Minkowski spacetime.
The Lagrangian for a particle in the Rindler spacetime is
$$
{\cal L} = -\mu_0 \sqrt{\big(1+gz(t)\big)^2 -  \sum_{i=x,y,z} {v^i (t)}^2 },
$$
where $\mu_0$, $z(t)$ and $v^i(t)$ are the rest mass, the position in $z$-direction and the velocity of the particle with respect to the time $t$.
The conjugate momentum to $x^i$ becomes
\begin{equation}
p^i = \frac{\mu_0 v^i}{\sqrt{ (1+gz)^2 - v^2}}
	= \frac{\mu_0 v^i \sqrt{1+p^2/\mu_0^2}}{ 1+gz},
\end{equation}
where $v^2= \sum_{i=x,y,z} {v^i (t)}^2$ and 
$p^2=\mu_0^2 v^2/\sqrt{(1+gz)^2 -v^2}$.
Now, the Hamiltonian is given by 
\begin{equation} \label{H}
{\cal H} \equiv \sum p_i v^i- {\cal L}
	= \mu_0 \Big(1+gz\Big) \sqrt{ 1+\frac{p^2}{\mu_0^2}}.
\end{equation}
Note that the Hamiltonian and momentum satisfies the required relation as a four-vector, $g^{00} H^2 + g^{ij} p_ip_j=-\mu_0^2$.

Now, let us consider a static distribution of fluid composed of $N$-noninteracting identical particles in a box in Rindler spacetime. 
In general, the system can be treated as a canonical ensemble of ideal gas.
Parts of the subject discussed here were studied in Ref.~\cite{LouisMartinez:2010nh} and its nonrelativistic version was given in Ref.~\cite{Landsberg:1994}.
Theoretical considerations on the matter states in the presence of gravity were presented in Refs.~\cite{Martinez:1996vy,Martinez:1996,Sorkin:1981}.
Let the height and the bottom area of the box be $2L$ and $A$, respectively.
Also, let the center of the box be located at $z=0$ and its $z$-axis be
anti-parallel to the gravity.  
As the height of the box is restricted to be $L<g^{-1}$, the bottom of the box does not stretch behind the event horizon.
The distribution of the particles should be independent of the Killing directions, $x$ and $y$. 
A concrete stress energy tensor description of the system of identical particles is given in Ref.~\cite{MTW}.
For a given stress energy tensor $T_{ab}$, the energy density for an observer having the timelike Killing vector $\chi^a= (\partial/\partial t)^a$ and the timelike unit normal $n_a = (1+gz)^{-1} \chi_a$ is defined by $E(x) =  T^a_{~b} \chi^b n_a$.
Explicitly, 
\begin{equation}\label{rho}
E(x^a) = -(1+gz) T^{0}_{~0}= (1+gz) \rho(x^a).
\end{equation}
The pressure can be obtained by integrating the Euler equation for stationary spacetime,
\begin{equation}
\frac{\partial P}{\partial x^0} =0, \qquad
\frac{\partial P}{\partial x^j} = - (\rho +P) \frac{\partial \log 
	\sqrt{-g_{00}}} {\partial x^j} \quad \Rightarrow \quad
	\Big(\frac1g+ z\Big) \frac{\partial P}{\partial z} = -  (\rho +P) 
	. 
\end{equation}
For a given $\rho(z)$, the pressure is determined to be
\begin{equation} \label{Euler}
P(z) = -\frac{g}{1+gz} \int^z dz' \rho(z'),
\end{equation}
where $\rho(z)$ and $P(z)$ are defined in an orthonormal coordinates system.

In the next section, Sec. II, we consider the statistical description of ideal gas in Rindler spacetime by using canonical ensemble. 
Then, in Sec. III, various approximations of the theory are investigated with respect to the values of temperature and gravity. 
In Sec. IV, we obtain the first law of thermodynamics and the EOS for each regime and investigate its consequences.
We summarize our results in Sec. V.

%================================
\section{Ideal gas in Rindler spacetime}
The system inside the box is assumed to be isolated from outside and to be in
contact with a thermal bath of temperature $T\equiv 1/(k_B\beta)$ at $z=0$.
With this, we treat the system as a canonical ensemble.  
A statistical description of the system of particles in Rindler spacetime was given in Ref.~\cite{LouisMartinez:2010nh} by incorporating grand canonical ensemble. 
Part of this section reproduces their results with a little difference in notation.

In the presence of a gravity, the distribution of particles are dependent on the position.
The number density of particles in a unit phase volume at $(x^i, p_j)$ is given by
\begin{equation}
n(z,p) = \frac{N}{h^3Z_1} \, 
	e^{-\beta H}
           = \frac{N}{h^3 Z_1} e^{-\beta \mu_0 (1+ gz) 
           	\sqrt{1+ p^2/\mu_0^2}}
         =\frac{N}{V \cZ}   \frac{1}{4\pi\mu_0^3}e^{-\beta \mu_0 (1+ gz) \sqrt{1+ p^2/\mu_0^2}}
         .
\end{equation} 
Here $Z_1= h^{-3} \int e^{-\beta H} d^3p d^3 r $ is the one-particle partition function and the gravity dependent normalization is encapsulated in the function $\cZ$ as 
\begin{equation}
 \cZ(\alpha_+, \alpha_-) = 
	\frac{1}{2X}  \left[ -\frac{K_1(\alpha_+)}{\alpha_+} 
		+\frac{K_1(\alpha_-)}{\alpha_-}    \right],
\qquad \alpha_\pm = \beta \mu_0 (1 \pm \G), \quad X \equiv \frac{\alpha_+-\alpha_-}2,
\end{equation} 
where $K_1(\alpha_\pm)$ is the modified Bessel function with the order 1 and
the argument $\alpha_\pm$, and $\G=gL/c^2$ represent the strength of the gravitational potential due to the external gravity $g$. 
$\beta = T^{-1} $ represents the global temperature measured at $z=0$.

Integrating over the momentum, one gets the number density per unit volume at $z$ as  
\begin{equation}
n(z) = \frac{N}{V \cZ} \frac{K_2(\alpha)}{ \alpha},
\label{density}
\end{equation}
where $\alpha = \beta \mu_0\left( 1+gz\right)$.  Here, we note the recurrence relation between $K_1(\alpha)$ and $K_2(\alpha)$ 
\begin{equation}
\frac{d}{d\alpha} \frac{K_1(\alpha)}{\alpha} = - \frac{K_2(\alpha)}{\alpha},
\end{equation}
which can be inferred from the general recurrence relations of modified Bessel functions for any order $\nu$ with an argument $x$ : 
\begin{equation} \label{BesselK}
K_{\nu-1}(x)-K_{\nu+1}(x) = -\frac{2\nu}{x} K_\nu(x), \qquad K_{\nu-1}(x) + K_{\nu+1}(x)= -2 K'_\nu(x).
\end{equation}
Integrating over the volume for $n(z)$ in Eq.(\ref{density}), one gets  
the normalization condition,
 $\int d^3r\, n(z) =N$.

The number density decreases with $z$ monotonically. 
When the value of $\alpha $ is allowed to be close to zero by any means (due to large $L$ or strong gravity), 
most of the particles will be gathered around a very narrow surface at $\alpha \, \approx \alpha_- = \mu_0g \delta/k_BT \approx 0$, because $n(z) \propto \alpha^{-3}$, where $\delta\equiv g^{-1}-L$ is the distance to the event horizon from the bottom of the box.
Therefore, the system behaves as if it is located on a two dimensional plane.
This phenomenon is closely related to the area law of blackhole entropy~\cite{Kolekar:2010py}. 
From Eq.~\eqref{rho}, we calculate the energy density at $z$, 
\begin{eqnarray}
\rho(z) \equiv \frac{1}{1+gz}\int d^3p  \,H(z,p) n(z,p)
 &=& -\frac{M_0}{V \cZ} \left(\frac{\partial}{\partial\alpha} 
 		\frac{K_2(\alpha)}{\alpha} \right) ,
\end{eqnarray}
where $M_0 \equiv N\mu_0$.
From Eq.~\eqref{Euler}, we get the pressure given by
\begin{equation}
P(z) = \frac{M_0}{V \cZ} \frac{K_2(\alpha)}{\alpha^2} = \frac{n(z) k_B T}{1+ gz}.
\end{equation}
Note that the pressure satisfies the ideal gas law $P(z)= n(z) k_B T(z)$ with the well-known local temperature $T(z) = T/(1+ gz)$.
The local temperature $T(z)$ describes the temperature of a system located at $z$ in equilibrium with the heat bath of temperature $T$. 
Note that the two temperatures are related by the redshift factor.
The pressure is huge for small $\alpha$,
$
P(z) \approx 2 M_0/V\cZ\alpha^4 .
$
The pressure difference from the top to the bottom is 
$$
\Delta P = \frac{M_0}{V \cZ} \left( \frac{K_2(\alpha_+)}{\alpha_+^2} - \frac{K_2(\alpha_-)}{\alpha_-^2}\right).
$$
The pressure average does not satisfy the EOS in Eq.~\eqref{EoS} because the temperature is position dependent. It is 
$$
P_{\rm avg} = \frac{1}{V} \int d^3r\, P(z) = \frac{M_0}{V\cZ} \frac{1}{2X}\int_{\alpha_-}^{\alpha_+} \frac{K_2(\alpha)}{\alpha^2}d\alpha 
=   \frac{M_0}{V\cZ} \frac{G(\alpha_+)-G(\alpha_-)}{2X} ,
$$  
where $G(\alpha)$ can be expressed as a Meijer G-function.

One may also define the pressure in Rindler frame as $p(z) \equiv (1+gz) P(z) = n(z) k_B T$ as was done in Ref.~\cite{LouisMartinez:2010nh}.
The Rindler pressure satisfies
\begin{equation} \label{pavg}
p_{\rm avg} = \frac{1}{V} \int d^3r\, p(z) =\frac{Nk_B T}{V}, \qquad
\Delta p = - 2 [u(\alpha_+,\alpha_-) +1] X p_{\rm avg},
\end{equation}
where
\begin{equation}
u(\alpha_+,\alpha_-)     \equiv \frac{1}{2 X \cZ} \Big( \frac{K_2(\alpha_-)}{\alpha_-} -
 	\frac{K_2(\alpha_+)}{\alpha_+} \Big) -1 .
\label{u}	
\end{equation}
Note here that the $u(\alpha_+,\alpha_-)$-term vanishes in the Newtonian limit, reproducing the result in Ref.~\cite{Kim:2016txr}. 

The $N$-particle partition function in the canonical ensemble is given by  
$\log Z_N = N \log Z_1 - \log N !,$ whose logarithm is
\be{ZN}
\log Z_N[V,\beta, \G] 
=  N\log \frac{V}{(\hbar/\mu_0)^3N}+ N
	+\frac{3N}2 \log \frac{k_B T}{2\pi \mu_0} 
+ N\log \left[ \frac{(\beta \mu_0)^{3/2} }{\sqrt{\pi/2}}\, 
	\cZ  \right] ,
\ee
where the volume $V$ is divided by a natural volume element, $(\hbar/\mu_0c)^3$. 
We have restored $c$ here to show the dimension as a volume clearly.

The energy and entropy of the $N$-particle system in Rindler spacetime are given by
\begin{eqnarray} 
M_R(T,\G) &\equiv& -\left(\frac{\partial \log Z_N}{\partial \beta}\right)_V 
	= Nk_B T \,m(\alpha_+,\alpha_-) 
	, \label{MR}\\
\frac{S_N(V,T,\G)}{Nk_B} 
&\equiv& \frac{M_R}{N k_B T} + N^{-1} \log Z_N 
=  \log\frac{e V/N}{2\pi^2(\hbar/\mu_0)^3} + s(\alpha_+,\alpha_-),
\label{SN}
\end{eqnarray}
where  
\begin{equation} \label{m,s}
m(\alpha_+,\alpha_-)\equiv 1-  \frac{K_2(\alpha_+)
		-K_2(\alpha_-)}{ 2X \cZ } , \qquad
s(\alpha_+,\alpha_-)\equiv  m(\alpha_+,\alpha_-)+ \log \cZ.\end{equation}
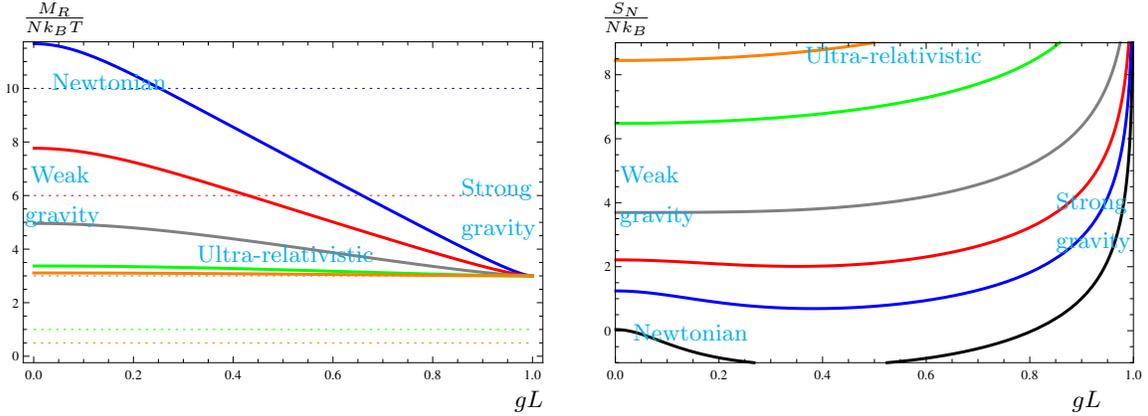
\begin{figure}[ht]
\begin{center}
\begin{tabular}{cc}
\includegraphics[width=.4\linewidth,origin=tl]{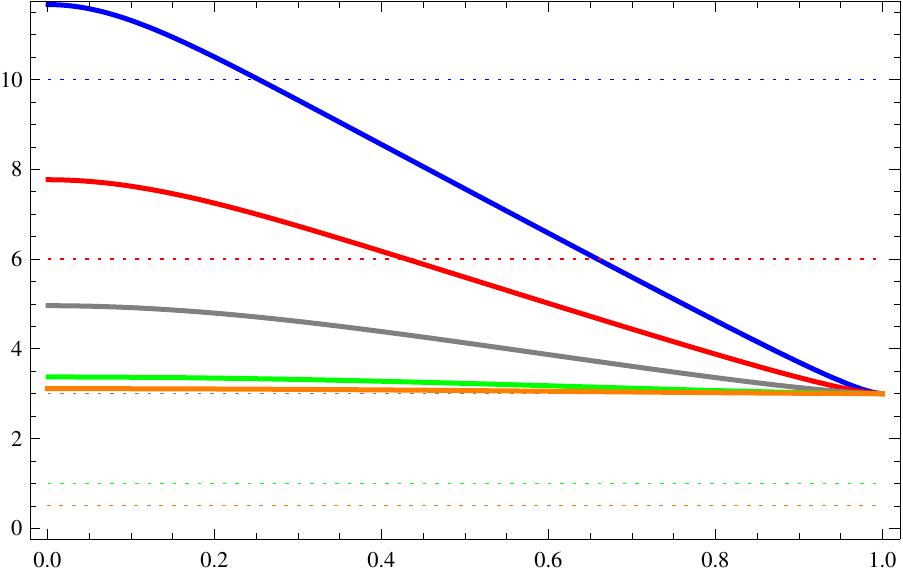}&
\qquad
\includegraphics[width=.4\linewidth,origin=tl]{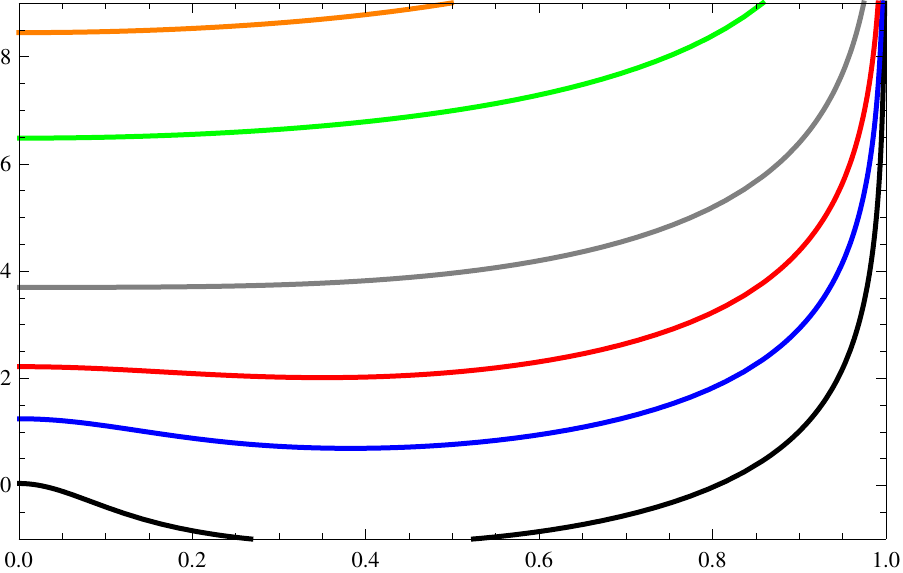}
\end{tabular}
\put (-207,75) {$\frac{S_N}{N k_B}$  }
\put (-29,-70) {$gL$  }
\put (-429,75) { $\frac{M_R}{Nk_B T}$}
\put (-240,-70) {$gL$  }
\textcolor{cyan}{
\put(-415,50){Newtonian}
\put(-360,-15){Ultra-relativistic}
\put(-423,15){Weak} 
\put(-425,0) {gravity}
\put(-260,10){Strong} 
\put(-260,-5){gravity}
\put(-195,-45){Newtonian}
\put(-130,60){Ultra-relativistic}
\put(-200,15){Weak} 
\put(-200,0) {gravity}
\put(-35,5){Strong} 
\put(-35,-10){gravity}}
\end{center}
\caption{The total energy $M_R(\beta \mu_0, X)$ (Left) and entropy $S_N(\beta \mu_0, X)$ (Right).
The energy is plotted for $\beta \mu_0 = 10, 6, 3, 1$ and $1/2$, respectively from above.  
The dotted line denotes the rest mass $\mu_0$ of the corresponding color.
In the right picture, the entropy is plotted for $\beta \mu_0 = 20,10, 6, 3, 1$ and $1/2$, respectively from the bottom.  
We also choose $V/(\hbar/\mu_0)^3/N = 100$.
}
\label{fig:P}
\end{figure}
As shown in Fig.~\ref{fig:P}, the energy monotonically decreases with $\G$ representing the attractiveness of the gravity. 
However, it is larger than $3Nk_B T$ for all $\G$.
This result is critically different from  
that of the Newtonian gravity where negative energy state is allowed because of the negative gravitational potential energy.  
As for the entropy, it decreases with $\G$ due to the ordering effect of gravity in the Newtonian regime where the temperature is low and gravity is weak.
However, when the gravity is strong or the temperature is high, the situation changes significantly. 
The entropy increases with the gravity and even diverges at $\G=1$.
We discuss in the next section, Sec. III, each specific regime denoted by ``Newtonian", ``Weak gravity", ``Ultra-relativistic" and ``Strong gravity" in Fig.\ref{fig:P}.

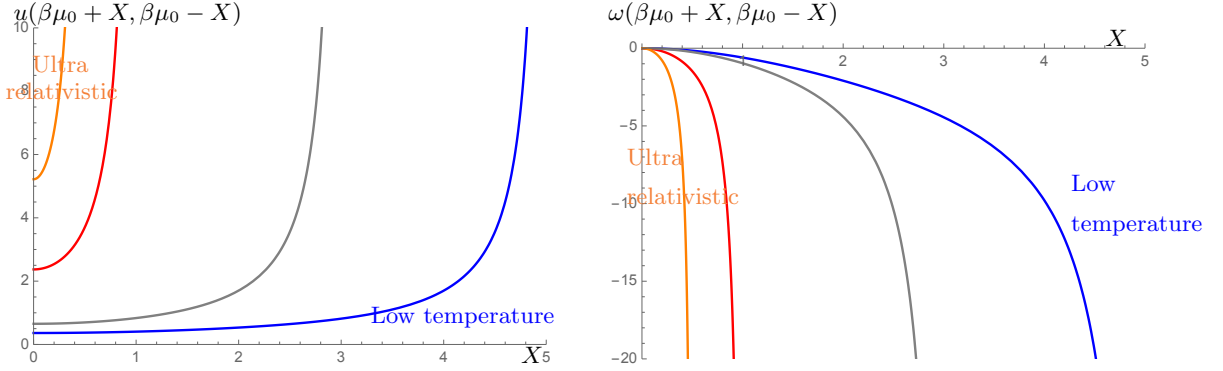
\begin{figure}[hbt]
\begin{center}
\begin{tabular}{cc}
\includegraphics[width=.4\linewidth,origin=tl]{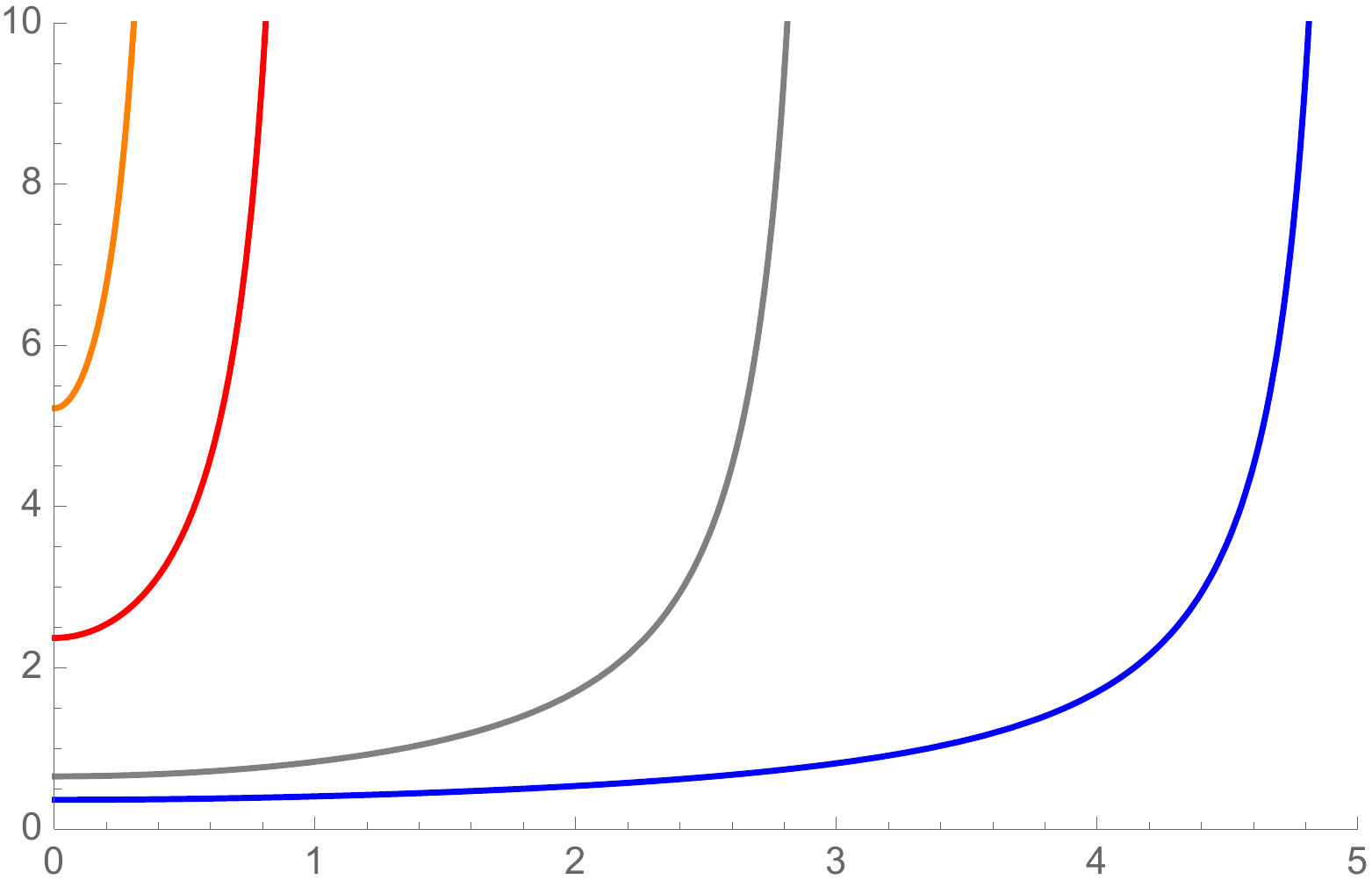}&
\qquad
\includegraphics[width=.4\linewidth,origin=tl]{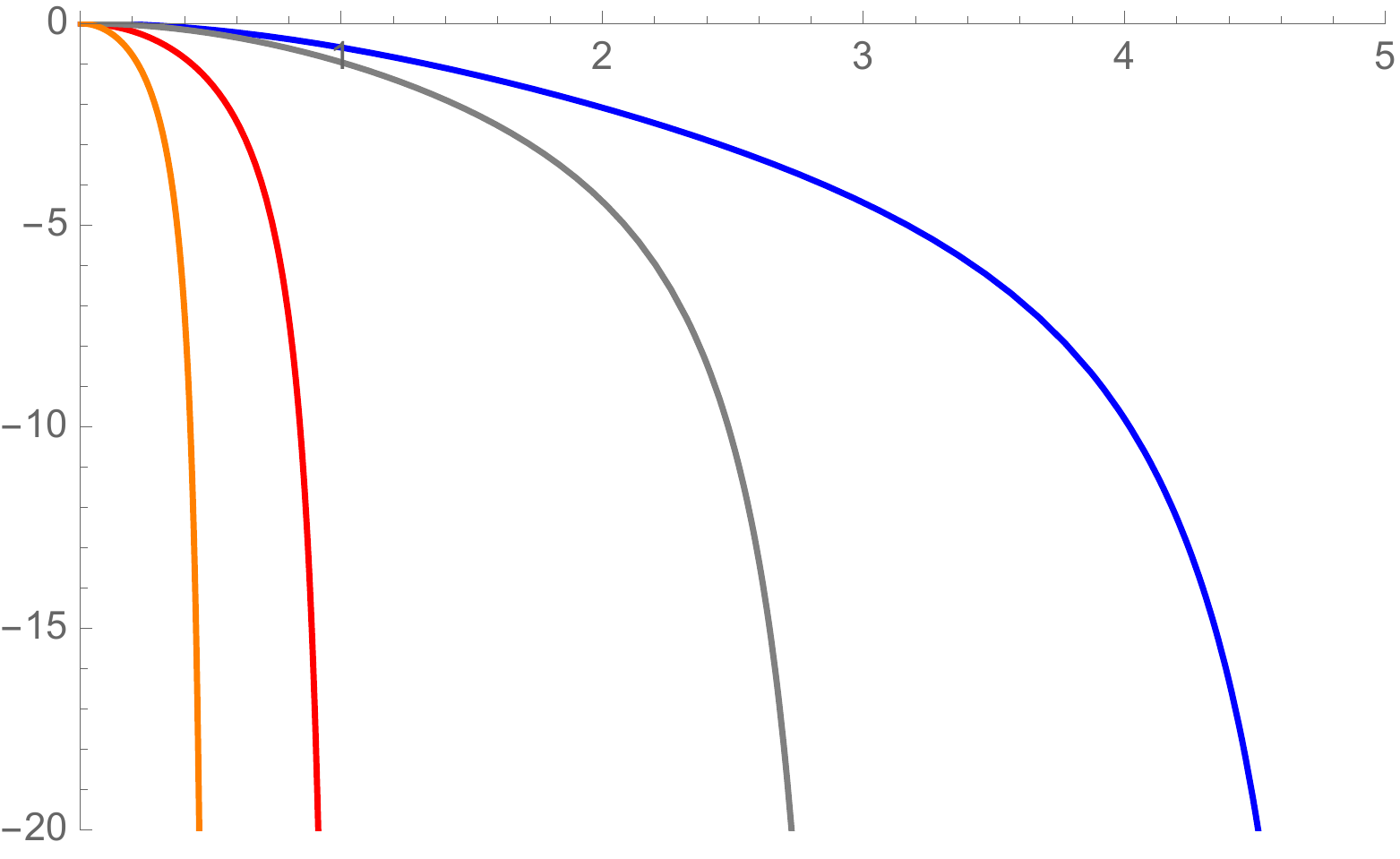}
%\\
%(a) %& (b)
\end{tabular}
\put (-207,70) {$\omega(\beta\mu_0 +X,\beta\mu_0-X)$  }
\put (-19,60) {$X$  }
\put (-435,70) { $u(\beta\mu_0 +X,\beta\mu_0-X)$}
\put (-240,-60) {$X$  }
\textcolor{Orange}{
\put(-425,50){Ultra}
\put(-435,40){relativistic}
\put(-200,15){Ultra} 
\put(-200,0) {relativistic}
}
\textcolor{blue}{
\put(-300,-45){Low temperature} 
\put(-35,5){Low} 
\put(-35,-10){temperature}}
\end{center}
\caption{The internal energy and the gravitational potential energy.
Here, $\beta \mu_0 = 1/2, 1, 3,$ and $ 5$, respectively from the left.
}
\label{fig:uom}
\end{figure}

The total energy~\eqref{MR} can also be obtained from the formula $M_R=\int T^a_{~b} \chi^b d\Sigma_a= \int T^a_b \chi^b n_a d^3 r= A\int dz \rho(z) (1+gz)$. 
The internal energy is given by the integration of the difference between the density and the rest mass over the volume,
\begin{equation} \label{Uint}
U_{\rm int}
    \equiv \int d^3r \left[ \rho(z) - \mu_0 n(z) \right] 
     =M_0u(\alpha_+,\alpha_-) .
\end{equation}
The internal energy diverges as $gL \to 1$ as shown in the left panel of Fig.~\ref{fig:uom}.
The gravitational potential energy is given by subtracting the rest-mass energy+the internal energy from the total energy:  
\begin{equation}\label{Omega}
\Omega \equiv M_R - U_{\rm int} - M_0
	= Nk_BT \, \omega(\alpha_+,\alpha_-), 
\end{equation}
where
\begin{equation} \label{omega}
\omega(\alpha_+,\alpha_-)\equiv 1-\frac{1}{2 \cZ }
 	\left( \frac{K_2(\alpha_+)}{\alpha_+}+ 
	\frac{K_2(\alpha_-)}{\alpha_-} \right) 
\end{equation}
satisfies
\begin{equation}\label{omu}
\omega= m- \beta \mu_0 ( u+1).
\end{equation}
In the presence of a gravity, the gravitational potential energy is negative definite as in right panel of Fig.~\ref{fig:uom}.

\begin{figure}[bht]
\begin{center}
\begin{tabular}{cc}
\includegraphics[width=.4\linewidth,origin=tl]{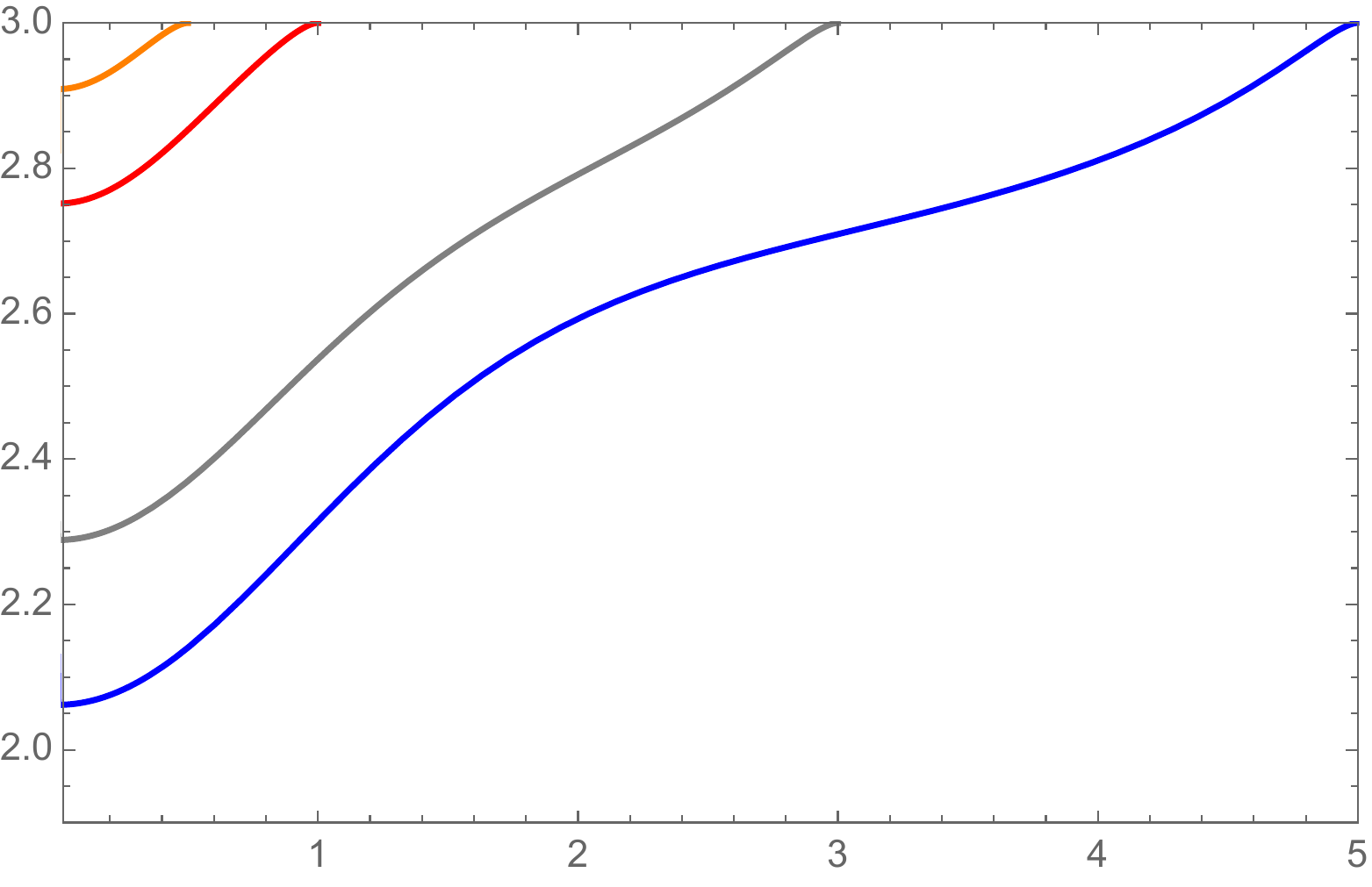}&
\qquad
\includegraphics[width=.4\linewidth,origin=tl]{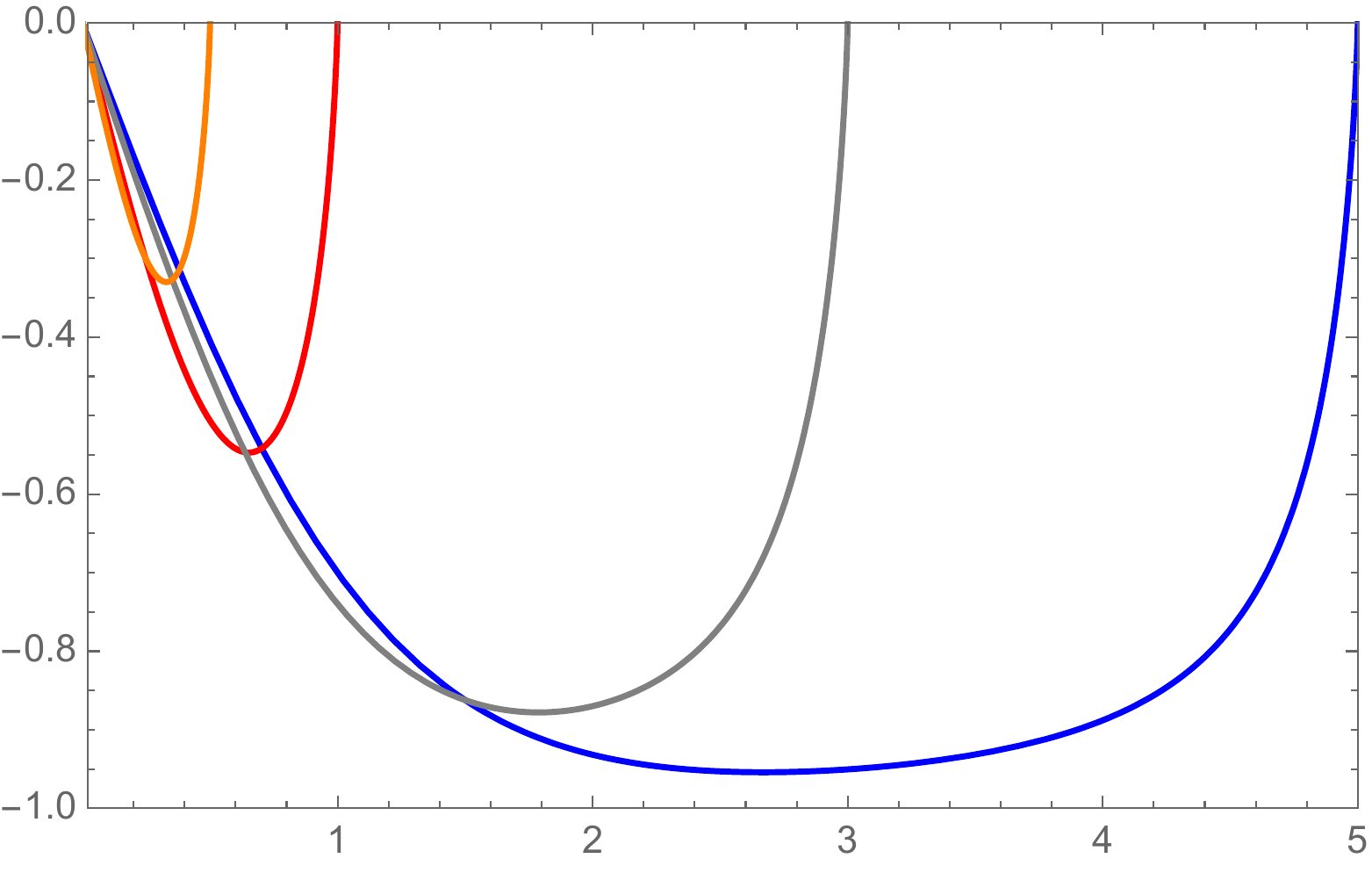}
%\\
%(a) %& (b)
\end{tabular}
\put (-207,70) {$C_T(\beta\mu_0 +X,\beta\mu_0-X)$  }
\put (-19,-60) {$X$  }
\put (-435,70) { $C_V(\beta\mu_0 +X,\beta\mu_0-X)$}
\put (-240,-60) {$X$  }
\textcolor{Orange}{
\put(-425,50){High}
\put(-435,40){temperature}
\put(-200,45){High} 
\put(-200,30) {temperature}}
\textcolor{blue}{
\put(-300,45){Low temperature} 
\put(-35,5){Low} 
\put(-35,-10){temperature}}
\end{center}
\caption{The heat capacity (L) and the gravity capacity (R).
Here, $\beta \mu_0 = 1/2, 1, 3,$ and $ 5$, respectively from the top.
}
\label{fig:HC}
\end{figure}
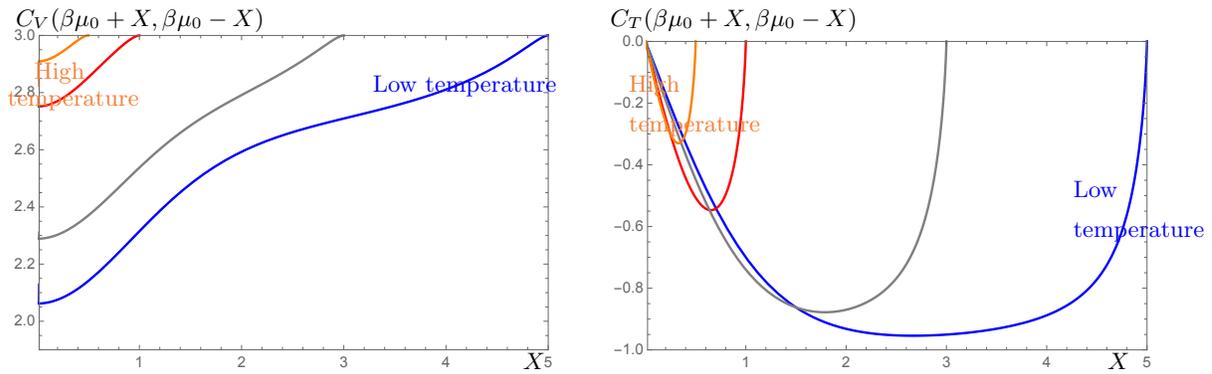

The heat capacity for fixed volume and gravity is
\begin{equation} \label{CV}
\frac{C_{V}}{Nk_B} 
\equiv \frac{1}{N k_B} \left(\frac{\partial M_R}{\partial T}\right)_{V,\G}
 = m(\alpha_+,\alpha_-)  
    - \mu_0 \beta \partial_S m(\alpha_+,\alpha_-) 
    - X \partial_A m(\alpha_+,\alpha_-) ,
\end{equation} 
where we use  
$(\beta \partial_\beta)_\G = 
	\alpha_+ \partial_{\alpha_+} 
	+ \alpha_- \partial_{\alpha_-}
= \frac{\alpha_++\alpha_-}2 \partial_S + \frac{\alpha_+-\alpha_-}2
	 \partial_A$, and introduce abbreviated notations
$$
\partial_S = \partial_{\alpha_+} + \partial_{\alpha_-}, 
\qquad 
\partial_A = \partial_{\alpha_+} - \partial_{\alpha_-}. 
$$
By using these notations, we find
\be{uom}
u(\alpha_+,\alpha_-) = -1-\partial_{S}  \log \cZ(\alpha_+,\alpha_-) , \qquad
\omega(\alpha_+,\alpha_-)
	= - X \partial_A \log \mathcal{Z}(\alpha_+,\alpha_-) .
\ee
As shown in Fig.~\ref{fig:HC}, the heat capacity $C_V$ monotonically increases to be $3$ at $X  = \beta\mu_0$.

Note that both the gravitational potential and the internal energy are dependent on the external gravity. 
The energy is also stored in the distribution of matters due to the gravity. 
The heat capacity for fixed temperature and volume is given by 
\begin{equation} \label{HC:T}
\frac{C_T}{M_0} \equiv M_0^{-1}\left(\frac{\partial  M_R}{\partial \G}\right)_{T,V} 
= \partial_A m 
	 ,
\end{equation}
where we use $(\partial_\G)_{T,V}= \beta \mu_0\partial_A$ and Eq.~\eqref{omu}.
We plotted both $C_V$ and $C_T$ in Fig.~\ref{fig:HC}.
Interestingly, $C_T$ goes to zero in the extreme gravity limit, $\G\to 1$.

%=============================
\section{Various regimes of the theory}
In this section, we consider the following four physically distinct regimes of approximations depending on the values of the parameters $\beta \mu_0$ and $\G$.  
They are the Newtonian regime ($\beta \mu_0 \gg 1$ with $\G<1$), the weak gravity regime ($\G\ll 1$), the ultra-relativistic regime (the average kinetic energy of particles is very high relative to its gravitational potential energy; $ \mu_0 gL \ll k_B T$), and the strong gravity regime ($\G \approx 1$). 
In the left panel of Fig.~\ref{fig:P},  
these four regimes correspond to the left side, the right side, the lower side and the upper left corner for the weak gravity regime, the strong gravity regime, the ultra-relativistic regime, and the Newtonian regime, respectively. 

In the Newtonian limit, one gets $M_R \approx  M_0 + 3Nk_B T/2$ as expected. 
For both of the ultra-relativistic regime 
and the strong gravity regime, the energy becomes $M_R \approx 3Nk_B T$. 
More details of each regime are discussed in the subsections below.
%

%============================
\subsection{The Newtonian regime}
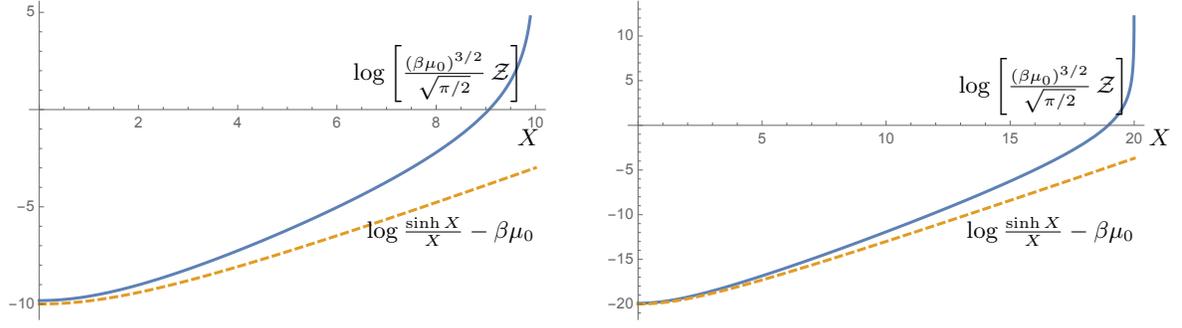
\begin{figure}[ht]
\begin{center}
\begin{tabular}{cc}
\includegraphics[width=.4\linewidth,origin=tl]{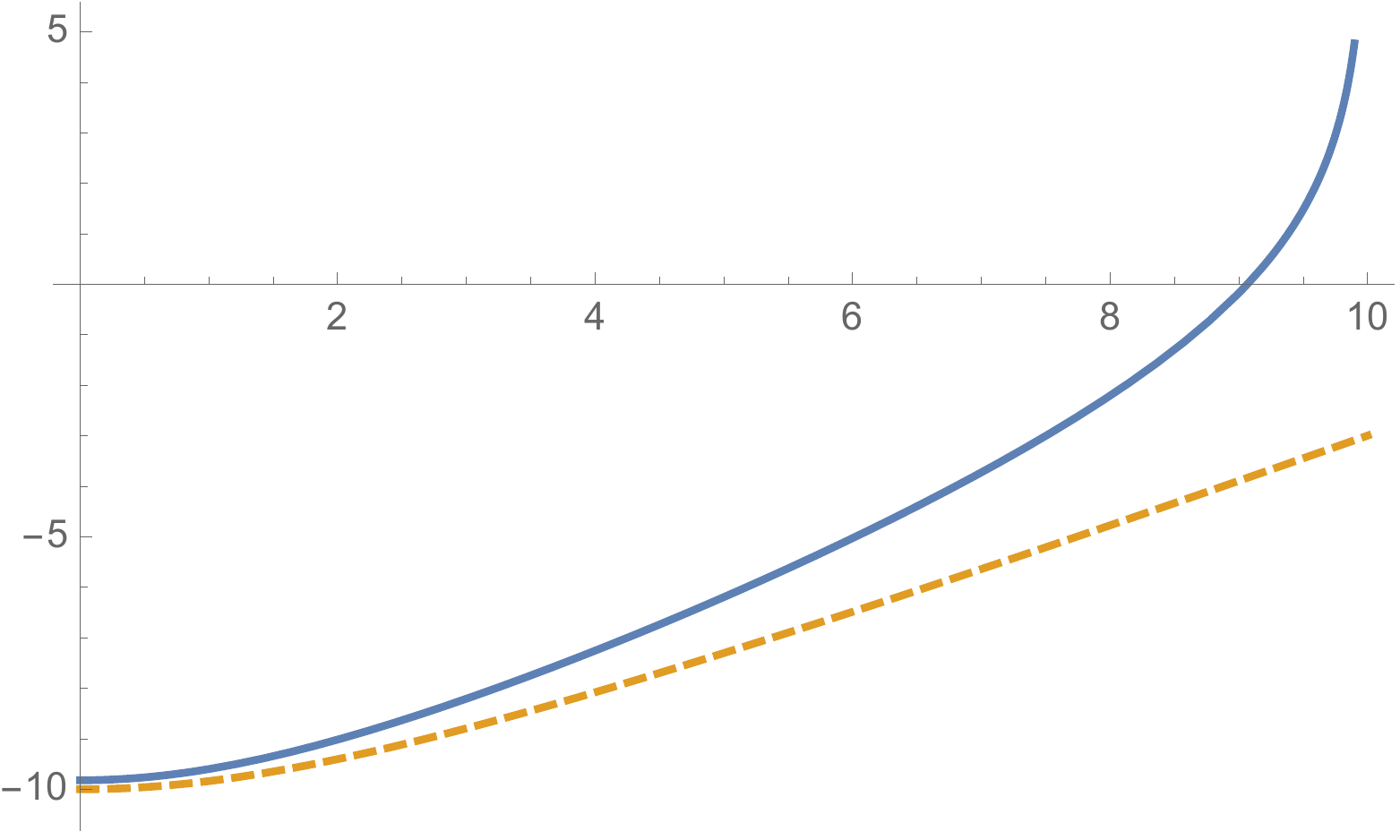}&
\qquad
\includegraphics[width=.4\linewidth,origin=tl]{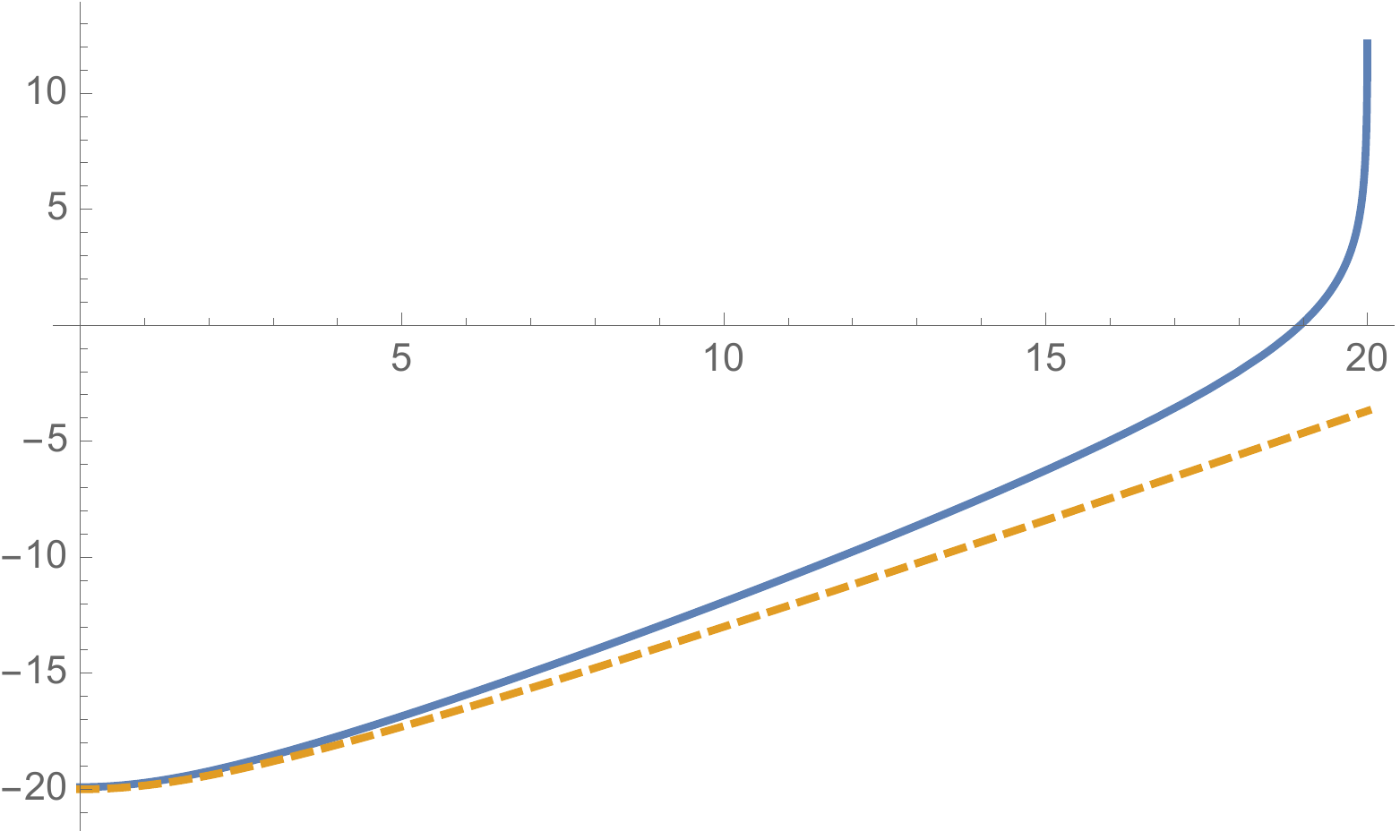}
%\\
%(a) %& (b)
\end{tabular}
\put (-73,30) {$\log \left[ \frac{(\beta \mu_0)^{3/2} }{\sqrt{\pi/2}}\, 
	\cZ  \right]$}
\put (-70,-25) {$\log \frac{\sinh X}{X}- \beta\mu_0 $  }
\put (-1,10) {$X$  }
\put (-302,35) {$\log \left[ \frac{(\beta \mu_0)^{3/2} }{\sqrt{\pi/2}}\, 
	\cZ  \right]$}
\put (-300,-25) { $\log \frac{\sinh X}{X}- \beta \mu_0 $}
\put (-240,10) {$X$  }
\end{center}
\caption{The partition functions for $\beta\mu_0=10$ (L) and  $20$ (R).
The blue and yellow dashed curves correspond to the last term of the logarithm of the exact partition function in Eq.~\eqref{ZN} and its Newtonian approximation.
}
\label{fig:znewtonian}
\end{figure}

In the Newtonian regime, the kinetic energy of a particle should be smaller than its rest mass, $\beta \mu_0 \gg 1$. 
The gravity should also be small, $gL\ll 1$, so that the time dilation effect is unimportant.
In the presence of these conditions, the Bessel function can be approximated to be
 $
K_n(x) \to \sqrt{\pi} e^{-x}/\sqrt{2x} 
$ 
as $x \to \infty$.
Thus, the partition function~\eqref{ZN} becomes
\begin{equation} \label{ZN:NR}
\log Z_N \to N\log \frac{V/N}{(\hbar/\mu_0)^3} +N
	+\frac{3N}2 \log \frac{k_B T}{2\pi \mu_0}  +N\left( \log \frac{\sinh X}{X}- \beta\mu_0\right) .
\end{equation}
The difference of Eq.~\eqref{ZN:NR} from the exact form of the partition function~\eqref{ZN} is given by the difference of the term inside the bracket and the last term of Eq.~\eqref{ZN}, which are plotted in Fig.~\ref{fig:znewtonian}. 
As seen in this figure, the difference increases with $X$. 
Especially around $X\sim \beta\mu_0$, the Newtonian approximation fails to hold in any way. 
A small gap exists even with $X=0$, which is the special relativistic correction. 
In Fig.~\ref{fig:znewtonian}, one may note that the better agreement between the approximate result (yellow dotted line)
and the exact result (blue solid line) is attained for $\beta \mu_0 = 20$ compare to $\beta \mu_0 = 10$.
This gap vanishes at zero temperature. 

Eq.~\eqref{ZN:NR} reproduces the result in Ref.~\cite{Kim:2016txr} up to the rest-mass contribution, $-N \beta \mu_0$, which is neglected in nonrelativistic mechanics.
The total energy~\eqref{MR} and entropy~\eqref{SN} become, 
$$
M_R \to M_0 + N k_B T\Big(\frac52-X \coth X\Big), 
\qquad
\frac{S_N}{N k_B} \to S_0
           + \log \big(\frac{ \sinh X}{X}\big)+1 - X\coth X. 
$$
Both reproduce Newtonian results in Ref.~\cite{Kim:2016txr} up to the rest mass contribution, where $S_0$ represents the entropy at $X=0$. 
A classical problem in this form of entropy is that it takes negative value for strong gravity. 
This problem is relieved in the general relativistic description of the ideal gas at the present work as can be seen in the results of the strong gravity regime 
below in the subsection~\ref{sec:D}.
Another resolution to this problem was given in Ref.~\cite{Landsberg:1994} by considering quantum statistics. 
The internal energy~\eqref{Uint} goes to the value of the kinetic energy, 
$
U_{\rm int}  \to 3Nk_B T/2.
$
The gravitational potential energy~\eqref{Omega} becomes
$
\Omega \to N k_B T(  1-X\coth X) .
$
One may also notice that the heat capacities attain the corresponding nonrelativistic value.
For example, the heat capacity $C_V$ takes the form,
\be{HC:NR}
\frac{C_V}{N k_B T} =  \frac52 - X^2 \csch^2 X 
	+ \frac{3 + 12 X^3 \coth X \, \csch^2 X}{4\beta \mu_0} + O\big((\beta\mu_0)^{-2}\big). 
\ee  
An interesting difference from the ordinary nonrelativistic result comes from the first order in $1/\beta\mu_0$. 
Note that $C_V/Nk_B T$ has a non-vanishing first order correction term $15/4\beta\mu_0$ even when the external gravity vanishes, $X=0$.
This term comes from the contribution of special relativistic effect.

%================================
\subsection{The weak gravity regime}
The weak gravity regime assumes that the gravitational potential energy of a particle due to the external gravity is negligible compared to the average thermal energy, $X\ll 1$ i.e.  $\mu_0 gL \ll k_B T$. 
The Bessel function is expanded as 
$K_n(\beta\mu_0 \pm X) 
\approx K_n(\beta\mu_0)  \pm X K'_n(\beta\mu_0)+\cdots$.
In Fig.~\ref{fig:P}, one finds that the entropy has a local maximum (minimum) at $X=0$ in the Newtonian (ultra-relativistic) regime. 
It is interesting to find out when the characteristic difference begins. 
The entropy~\eqref{SN} takes the form,
\be{S:WG}
\frac{S_N}{Nk_B} = 
\log\frac{V/N}{2\pi^2(\hbar/\mu_0)^3} + \log \frac{K_2}{b} +
	\frac{b K_3}{K_2} +\frac12\left(\frac23+ \frac{4}{b^2} 
	- \left(\frac{K_1}{K_2}\right)^2 
	- \frac{3 K_1}{b K_2} \right) X^2+\cdots ,
\ee
where the arguments of the Bessel functions is $b\equiv\beta \mu_0$. 
Examining the coefficient of $X^2$, we find that it happens at $\beta\mu_0 = b_c \approx 3.13145$.  

Most of physical quantities do not have $O(X)$ contributions.
For example, the total energy~\eqref{MR} behaves as
$$
\frac{M_R}{Nk_B T} = \Big(\frac{\beta\mu_0 K_3}{K_2}-1\Big) 
	-\left(\frac{K_1K_3}{K_2^2} -\frac13\right) \frac{X^2}{2} +\cdots.
$$
Examining the zeroth order term, $M_R$ changes from $M_0 + 3Nk_B T/2$ to $3Nk_B T$ with $T$.
The graph is convex as shown in Fig.~\ref{fig:P} because the coefficient of $X^2$ is always negative definite. 
The internal energy~\eqref{Uint} is 
$
U_{\rm int} = ( -1-\beta\mu_0 + \beta\mu_0K_3/K_2
	) N k_B T +O(X^2).
$
The gravitational potential energy~\eqref{Omega} vanishes to 
that order. 

%====================================
\subsection{Ultra-relativistic regime}
The ultra-relativistic limit corresponds to the case that the kinetic energy of a particle is much larger than its rest mass, $\beta \mu_0 \ll 1$. 
In this case, the argument of the Bessel function is very small, implying
$$
K_1(x) \approx \frac1{x} + \frac{x}4( 2 \log \frac{x}{2} + 2\gamma-1)+O(x^3), \qquad
K_{n>1}(x) \approx \frac{2^{n-1} \Gamma(n) }{x^n}\left[1-\frac{x^2}{4(n-1)}+ O(x^4)\right]. 
$$
Now, the total energy~\eqref{MR},
\begin{equation} \label{m:UR}
M_R\approx Nk_B T \left[ 3-  \frac{(1-\G)^2}{4\G} \log \frac{1-\G}{1+\G} \times (\beta \mu_0)^2 \right],
\end{equation}
is close to $3Nk_B T$ and correction to that begins at $O(\beta^2\mu_0^2)$. 
The entropy in Eq.~\eqref{SN},
$$
\frac{S_N}{Nk_B} \approx \log \frac{e V/N}{2\pi^2(\hbar/\mu_0)^3} +
	3\log \frac{e k_B T}{\mu_0} + \log \frac{2}{(1-\G)^2}
	+\frac{ (1-\G^2)^2\arctanh\G}{4\,\arctanh\G} 
		\Big(\frac{\mu_0}{k_B T}\Big)^2+\cdots,
$$
has high value because of the logarithmic term of the temperature, where $\arctanh(x)$ denotes the inverse function of the hyperbolic function $\tanh (y)$.
The entropy increases with the external gravity $\G$.
Therefore, the negative entropy problem posed in the Newtonian regime disappears as previously stated. 
The internal energy~\eqref{Uint} and the gravitational potential energy~\eqref{Omega} are given by
\begin{equation} \label{UintOmega}
U_{\rm int} \approx N k_B T \frac{3+\G^2}{1-\G^2} -M_0 +O(\beta \mu_0), \qquad 
\Omega \approx -Nk_B T\frac{4\G^2}{1-\G^2}+ O\big(\beta \mu_0\big)     ,
\end{equation}
respectively. 
Both of the values diverge at $\G=1$ but their sum is finite, $3Nk_BT - M_0$, and independent of the external gravity. 

%====================================
\subsection{Strong gravity regime} \label{sec:D}
Contrary to the Newtonian gravity, the value of $\G$ does not grow indefinitely but is bounded above at $\G=gL/c^2 =1$, 
which corresponds to the case that the bottom of the box touches the event horizon of the Rindler spacetime. 
Therefore, the strong gravity limit is given by the condition $\G \approx 1$. 
At the present case, it is convenient to write the result down by using
\be{delta}
1-\G = 1- gL= g(g^{-1}-L) = g \delta. 
\ee
The total energy in Eq.~\eqref{MR} becomes 
\begin{equation}
M_R\approx Nk_B T \Big[ 3+  \Big(-\log \frac{\mu_0 g \delta}{2k_B T}
	 - \gamma -K_0(2\beta\mu_0)
	\Big) (\beta \mu_0)^2\times 
		(g \delta)^2 \Big]. 
\end{equation}
Note that most of the particles are located at the bottom of the box. 
Therefore, one may write the particle number $N$ in terms of surface number density as $n_S \equiv N/A$. 
Now, by using $V= 2AL\approx 2A g^{-1}$, the entropy~\eqref{SN} takes the form,
$$
S_N \approx A \, n_S \left[\log \left(\frac{e^4}{2\pi^2 (\hbar/\mu_0)^3 g n_S}\right)
	-3 \log \frac{\mu_0}{k_B T} 
	-2 \log (g\delta )+\cdots \right].
$$
The proportionality to the area becomes evident. 
It diverges as $\delta \to 0$. 

The internal energy~\eqref{Uint} and the gravitational potential energy~\eqref{Omega} become
\begin{equation} \label{UOmega:SG}
\frac{U_{\rm int}}{N k_B T} \approx  \frac{2}{g\delta} -\frac{\mu_0}{k_B T} + O(\delta), 
\qquad
\frac{\Omega}{N k_B T} \approx -\frac{2}{g\delta} + 3+ O(\delta).
\end{equation}
Both diverges as $\delta \to 0$ and sum of the two takes a finite value, $3Nk_B T -M_0$.

%============================
\section{Equation of state for adiabatic system}
In this section, we derive the EOS of the ideal gas system including the effect of external gravity.
As stated in the previous section, the ideal gas law is satisfied locally at each point, $P(z)= n(z) k_B T(z)$. 
The average value also satisfies $p_{\rm avg} = N k_B T/V$. 
Note however that $p_{\rm avg}$ is not the average of $P(z)$ but the average of $p(z) = (1+gz)P(z)$.
In this sense, the ideal gas law is not satisfied strictly. 

To see the EOS of an adiabatic ideal gas, let us begin with the derivation of the first law of thermodynamics. 
In canonical ensemble, the entropy is given by 
$S_N/k_B \equiv M_R/k_B T +\log Z_N$.
Differentiating both sides, we get 
\begin{equation} \label{dU}
dM_R = k_B T dS_N + M_R \frac{dT}{T}- k_B T d\log Z_N. 
\end{equation}
By explicitly differentiating $\log Z_N$ in Eq.~\eqref{ZN}, one gets
$$
d\log Z_N = \frac{M_R}{k_B T} \frac{d T}{T} + \frac{N}{V} dV 
	-N\omega(\alpha_+,\alpha_-)\, d\log\G. 
$$
Combining this with Eq.~\eqref{dU}, the first law of thermodynamics becomes 
\begin{equation}\label{1stlaw}
dM_R = k_B T dS_N - \frac{Nk_B T}{V} dV 
	+\Omega\, d\log \G .
\end{equation}
As seen in this equation, the contribution of the gravitational potential energy is included naturally in the first law similarly to the case of the nonrelativistic calculation in Ref.~\cite{Kim:2016txr}.
One may also write the second work term as $p_{\rm avg} dV$ by using Eq.~\eqref{pavg}. 

Now, we derive the EOS for an isentropic process, $dS=0$.
The first law of thermodynamics~\eqref{1stlaw} becomes $dM_R = -N k_B T d\log V + \Omega d\log \G$. 
Differentiating the energy in Eq.~\eqref{MR}, one gets
\begin{equation}
dM_R =  C_V dT + C_T d\G= 
	N k_B T \left[ \frac{C_V}{N k_B} d\log (Nk_BT) +
	(X \partial_A m)\, d\log \G \right].
\end{equation}
Equating the two equations,  we obtain
\begin{equation} \label{isentropic}
\frac{C_V}{N k_B} d\log (N k_B T)+ \frac{dV}{V} 
=\mathcal{C}_\G \,d\log \G; \qquad
\mathcal{C}_\G \equiv\omega - X \partial_A m .
\end{equation}
Further by using  $d\log \G =
d\log X + d\log (N k_B T)$ which comes from $\G= N k_B T X/(N\mu_0)$, we arrive at 
\begin{equation} \label{PVX}
\frac{dV}{V} 
=\CG\, d\log X- \mathcal{C}\,d\log (N k_B T) ,
\end{equation}
where
\begin{equation}
\begin{split}
\mathcal{C} \equiv \frac{C_V}{N k_B}-\omega + X \partial_A m
&= 	\beta \mu_0 (u+1 -\partial_S m) .
\end{split}
\end{equation}
Here, we use Eq.~\eqref{omu}. 
Regarding $V$ as a function of $T$ and $X$ only, the integrability condition of Eq.~\eqref{PVX},
\begin{eqnarray}
\left(\frac{\partial C_\G}{\partial \log Nk_B T}\right)_X 
= - (\beta \mu_0 \partial _S )(X\partial_A) [  \log \cZ +m]= 
-\left(\frac{\partial \mathcal{C}}{\partial \log X} \right)_T,
\end{eqnarray}
is satisfied, where we use Eq.~\eqref{uom}
and the commutativity of $(\beta\mu_0\partial_S)$ and $(X\partial_A)$. 

\begin{figure}[ht]
\begin{center}
\begin{tabular}{cc}
\includegraphics[width=.4\linewidth,origin=tl]{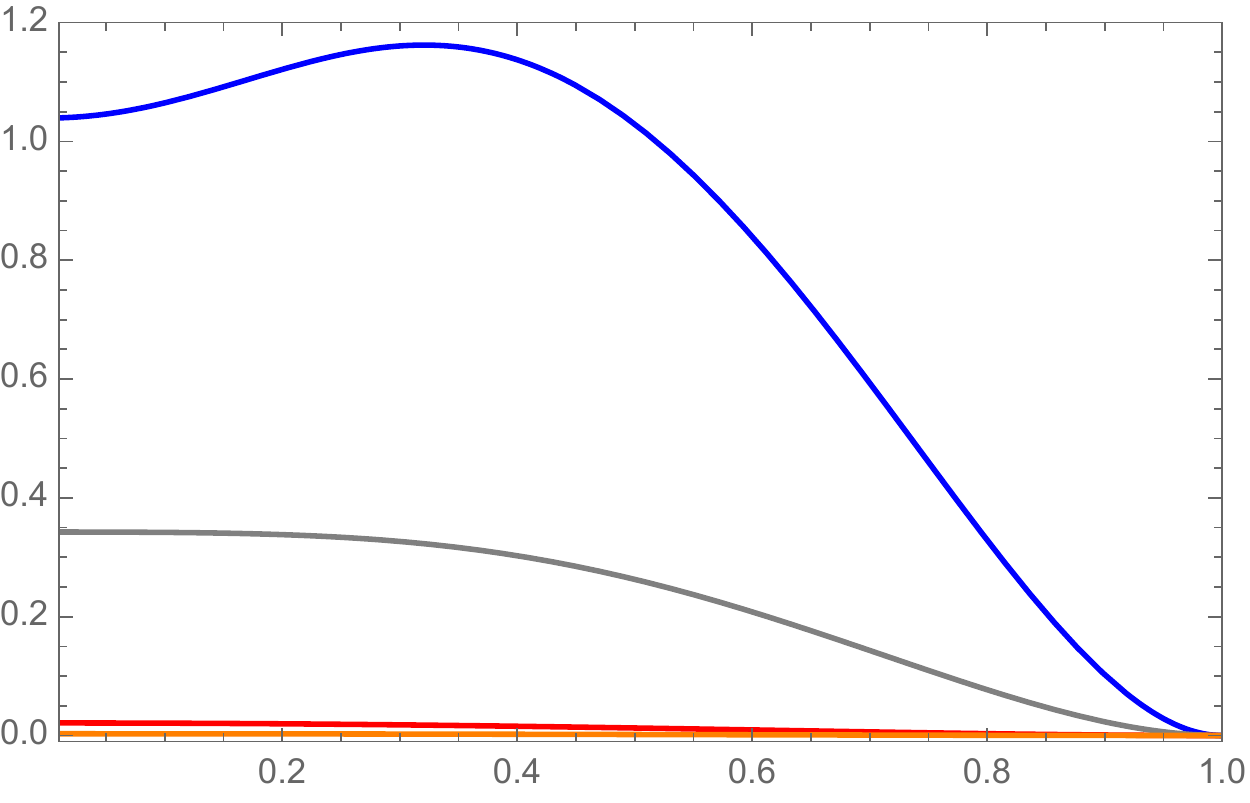}
\end{tabular}
\put (-237,55) {$ \frac{V(\alpha_+,\alpha_-)}K$}
\put (-20,-60) {$\G$  }
\end{center}
\caption{The EOS function for $\beta\mu_0 = 1/2, 1, 3,$ and $5$, respectively from the bottom.
}
\label{fig:eos}
\end{figure}

With these analysis, we can integrate Eq.~\eqref{PVX} exactly to give the EOS,
\begin{equation} \label{EoS:GR}
V =\tilde{K}\,\frac{e^{-m(\alpha_+,\alpha_-)}}{\cZ(\alpha_+,\alpha_-)}  .
\end{equation}
where ${\tilde K}$ is an integration constant having the dimension of volume.
The functional behavior of $V$ with respect to the change of the temperature and gravity is plotted in Fig.~\ref{fig:eos}.
At $X=0$, $(\partial V/\partial \G)_T =0$.
In the Newtonian limit, the volume increases temporarily and then decreases with $\G$. 
On the other hand, for high temperature with $\beta\mu_0\leq b_c$, the volume monotonically decreases with $\G$. 

%=============================
\subsection{The Newtonian gravity limit}
Let us first begin with the Newtonian gravity regime defined by 
$ \beta\mu_0 \gg 1$.
Eq.~\eqref{EoS:GR} becomes
\be{EoS:NR}
V = \Big(\sqrt{\frac8 \pi}\frac{ \tilde K}{e^{3/2}} \Big) \left( \frac{M_0}{Nk_B T}\right)^{3/2} \frac{X}{ \sinh X} \, e^{X\coth X-1 } + O(\beta\mu_0)^{1/2}.
\ee
For weak gravity, we have $P_{\rm avg} = p_{\rm avg}$ practically.
The relative difference,
$
(P_{\rm avg} -  p_{\rm avg})/p_{\rm avg} \approx -\Omega/M_0,
$ 
is of $O(k_B T/\mu_0 \times X^2)$.
Therefore, by using the prescription $N k_B T\to P_{\rm avg} V$, one can show that Eq.~\eqref{EoS:NR} reproduces the result obtained in Ref.~\cite{Kim:2016txr} for Newtonian gravity, i.e. $P =k_{\rm poly}\rho^\gamma$ with $k_{\rm poly} = \sqrt{\frac8 \pi}\frac{\tilde K}{e^{3/2}}$
and $\gamma = 3/2$.

%================================
\subsection{Weak gravity regime}
We next consider the weak gravity regime.
Eq.~\eqref{EoS:GR} becomes
\be{EoS:WG}
V = (e{\tilde K}) \frac{\beta \mu_0}{K_2(\beta\mu_0)} \exp\left[- \frac{K_3(\beta\mu_0)}{K_2(\beta \mu_0)}\right] \left[1-
  \left( \frac{2}{3}
	+\frac{4}{(\beta\mu_0)^2}
	-\frac{K_1^2}{K_2^2}
	- \frac{3}{\beta\mu_0} \frac{K_1}{ K_2} \right) X^2 +
	O(X^4)\right]. 
\ee
The special relativistic effect was given in Ref.~\cite{Karsch:1980}.
Let us observe the nonrelativistic and ultra-relativistic limits.
In the nonrelativistic limit, Eq.~\eqref{EoS:WG} becomes
$$
V \to \frac{\tilde K}{e^{3/2}} \sqrt{\frac2\pi} (\beta \mu_0)^{3/2}.
$$
This reproduces the $X\to 0$ limit of Eq.~\eqref{EoS:NR}.
In the ultra-relativistic limit with $\beta\mu_0 \to 0$, Eq.~\eqref{EoS:WG} becomes
$$
V \to \frac{{\tilde K}}{2 e^{3}}  (\beta \mu_0)^{3}.
$$
Therefore, the EOS changes drastically from the nonrelativistic one.
Note that the coefficient of the $O(X^2)$ term changes sign exactly at the value $\beta\mu_{0}= b_c$ as the characteristic change of the entropy happens in Eq.~\eqref{S:WG}.

%========================
\subsection{Ultra-relativistic regime}
In the ultra-relativistic regime, $\beta \mu\ll1$, the EOS in Eq.~\eqref{EoS:GR} becomes 
\begin{equation} \label{PVG2}
V = \frac{\tilde{K}}{2e^3} (1-\G^2)^{2}(\beta \mu_0)^{3} +O(\beta \mu_0)^5.
\end{equation}
Note that this relation does not provide directly the relation between the pressure and the energy density because $V$ does not inversely proportional to the energy density in the presence of a strong gravity.  
At the present limit, $M_R \approx 3Nk_BT = 3 p_{\rm avg}V$ giving $p_{\rm avg} = \rho_{\rm Rin}/3$ with $\rho_{\rm Rin}=M_R/V$ irrespective of the strength of external gravity. 
In this sense, at high temperature, any system of particles behaves as if it is a radiation from the point of view of a Rindler observer. 

%===============================
\subsection{Strong gravity limit}
In the strong gravity limit, $\G \approx 1$, to the first non-vanishing order, Eq.~\eqref{EoS:GR} becomes 
\begin{equation} \label{EoS:SG}
V \approx \frac{2\tilde{K}}{e^3} \delta^2 g^2 (\beta \mu_0)^3 \quad
	\Longrightarrow \quad 
	k_B T = \left(\frac{\mu_0\tilde{K}^{1/3}}{e}\right) \left( \frac{\delta^2} {A}\right)^{1/3}  g,
\end{equation}
where we use
$
V = 2AL \approx 2A g^{-1} .
$
Restoring $c$ after introducing dimensionless parameters $\bar K\equiv  \tilde{K}/(\hbar/\mu_0 c)^3$ and $\bar A\equiv  A /\delta^2$, we rewrite Eq.~\eqref{EoS:SG} as
\be{T:Unruh}
k_B T = \frac{2\pi}{e}\Big(\frac{\bar K}{\bar A}\Big)^{1/3} \times \left(\frac{\hbar g}{2\pi c}\right). 
\ee
The EOS indicates that the temperature of the system should be proportional to the external gravity at the center of the box. 
Up to the dimensionless part, the Unruh temperature~\cite{Unruh} for the Rindler spacetime arises naturally.
At the present situation, we can not identify whether the proportionality to the acceleration is a coincidence or not.
Usually, the Unruh temperature is expected in an accelerating frame due to the difference of the vacuum state from that of the Minkowski one, which characteristic must be a quantum mechanical nature. 
A similar situation happened at 1973. 
When studying a black hole, the four law of black hole mechanics were found~\cite{Bardeen:1973gs} through the classical theory of gravity.
From the resemblance to the laws of thermodynamics, physicists conjectured the existence of a black hole temperature up to a proportionality constant, which was determined later by Hawking~\cite{Hawking:1974sw} through the calculation of Hawking radiation of a quantum field around the black hole spacetime.
Based on the experience, the present result can also be an example expressing the close relation between the event horizon and thermodynamics. 
A similar prediction for the Unruh effect in classical field theory was also given by Higuchi and Matsas~\cite{Higuchi:1993fn}.

Recall that Eq.~\eqref{T:Unruh} is the re-writing of the strong gravity limit of the EOS, a relation between macroscopic physical parameters. 
It does not specify the temperature to a constant value but contains other macroscopic parameter than the temperature $T$.
Note that the particles are gathered at the bottom of the box located at $z=-L=-g^{-1} + \delta $. 
As a result, the system behaves as if it is a two-dimensional surface on the $(x,y)$-plane. 
In a sense, one may regard the system to be a localized system at the surface rather than an extended one from $-L$ to $L$.
Consequently, the normalized area $\bar A$ rather than the volume appears in the EOS~\eqref{T:Unruh}.  
It is helpful to see the EOS~\eqref{T:Unruh} in terms of the local temperature. 
We find that the EOS becomes independent of the gravity and takes the form of an ordinary two-dimensional EOS relating the area $\bar A$ and the temperature $T_{\rm local}$: 
$$
T_{\rm local} = \frac{T}{1+gz} = \frac{T}{g\delta} = \frac{\hbar}{eck_B \delta}\Big(\frac{\bar K}{\bar A}\Big)^{1/3} . 
$$
The value diverges in the $\delta \to 0$ limit, which is the effect of the redshift factor.
%============================
\section{Summary and Discussions}
 
In this work, we studied 
how  a strong external gravity affects on the equation of state (EOS) of matters.
To do this we have employed a canonical ensemble of classical monatomic ideal gas inside a box located in a Rindler spacetime. 
We, then, calculated physical parameters such as the total energy, the entropy, the heat capacity, and the gravitational potential energy. 
We also studied the local distribution of the particles and showed that the ideal gas law $P(z) = n(z) k_B T(z)$ is satisfied locally. 
However, this does not imply that the same relation is satisfied globally on the whole system in terms of the average values of pressure $P_{\rm avg}$ and temperature. 
 
Depending on the values of the kinetic energy to mass ratio and the strength of external gravity times height of the box, we analyzed the system in four physically distinct regimes of approximations, namely, the Newtonian regime, the weak gravity regime, the ultra-relativistic regime, and the strong gravity regime. 
In this work, the total energy is shown to decrease monotonically with the increase of the external gravity times the box height ($\G$) representing its attractiveness.
It is however shown to be higher than $3Nk_B T$ always, which is a crucial difference from the expectation of the Newtonian gravity.   
As for the entropy, it decreases with $\G$ in the Newtonian regime.
However, in the strong gravity regime and the ultra-relativistic regime, the entropy increases with the gravity.
This result can be a resolution of the negative entropy problem of Newtonian gravity. 

For an adiabatic system, we obtain a new EOS including the gravity contribution. Our new EOS reproduces the result in Ref.~\cite{Kim:2016txr} in the Newtonian regime.
In the strong gravity regime, the bottom of the box is very close to the event horizon of the Rindler spacetime. 
Because of the strong gravity, most particles shall be gathered at a narrow surface in the bottom of the box.
This results in the system of an effective two dimensional surface, which
is closely related with the area law of blackhole entropy.
A new result found in this work is that the temperature of the system is not a free parameter but tends to be fixed by the EOS to a value proportional to the external gravity, which is reminiscent of the Unruh temperature.   
From the point of view of a local observer located at the bottom of the box, the EOS is independent of the acceleration and should be described by local quantities such as the surface area of the system, the local temperature, and the distance to the horizon.
Writing the EOS from the point of view of a distant observer at the origin (center of the box), the proportionality of the temperature to the acceleration appears through the redshift factor.  
At present, we do not know whether the proportionality is a mere coincidence or not.

%\vspace{2cm}
\section*{Acknowledgment}
This work was supported by the National Research Foundation of Korea grants funded by the Korea government NRF-2013R1A1A2006548
and in part by the US Department of Energy grant DOE Contract No.~DE-FG02-03ER41260.
%%%%%%%%%%%%%%%%%%%%%
%\appendix
%\section{The pressure}

\end{document}